\documentstyle[prl,aps,epsf]{revtex}
\begin{document}

\twocolumn[
\hsize\textwidth\columnwidth\hsize\csname@twocolumnfalse\endcsname

\title{Quantum nondemolition measurements of a qubit } 

\author{D.V. Averin}

\address{Department of Physics and Astronomy, SUNY Stony Brook, 
Stony Brook, NY 11794-3800} 

\date{\today} 
 
\maketitle 

\begin{abstract}

The concept of quantum nondemolition (QND) measurement is extended to 
coherent oscillations in an individual two-state system. Such a 
measurement enables direct observation of intrinsic spectrum of these 
oscillations avoiding the detector-induced dephasing that affects the 
standard (non-QND) measurements. The suggested scheme can be realized 
in Josephson-junction qubits which combine flux and charge dynamics. 

\end{abstract} 
\pacs{PACS numbers: }

]

Quantum coherent oscillations in a two-state system (qubit) represent 
the most basic dynamic manifestation of quantum coherence between 
the qubit states. Motivated by potential application to quantum 
computation \cite{b1,b2} and conceptual interest in macroscopic 
quantum phenomena, significant effort is devoted at present to 
attempts to observe and study these oscillations in individual 
``mesoscopic'' qubits realized with Josephson-junction systems -- 
see, e.g., \onlinecite{b3,b4,b5,b6,b7}. One of the most direct ways 
of detecting the coherent oscillations in a qubit is to monitor 
them continuously with a weakly-coupled linear detector \cite{b8}. 
Spectral density of the detector output should exhibit then the 
spectral line at the oscillation frequency which contains 
information about the oscillation amplitude and decoherence rate, and 
has other interesting features. For example, the absolute intensity 
of the oscillation line demonstrates directly the quantum nature of 
the oscillations. It exceeds by a factor of two intensity of the 
classical harmonic oscillations of the same amplitude. Quantum 
mechanics makes larger intensity possible by combining harmonic 
oscillations of the probability with discrete jumps of the 
oscillating variable between the two states of the qubit \cite{b8}. 

The spectral line in the detector output, however, does not fully 
represent intrinsic spectral density of the oscillations. In the 
simplest measurement scheme the detector measures directly the 
oscillating coordinate, and thus tends to localize it, 
introducing extra dephasing in the dynamics of the oscillations. 
Such a ``backaction dephasing'' creates a fundamental limit, equal 
to 4, for the signal-to-noise ratio of the measurement, i.e., the 
ratio of the height of the oscillation line to the output noise of 
the detector. This limitation makes direct measurement of the quantum 
coherent oscillations in an individual qubit difficult, and leads 
to an interesting question whether a variant of quantum nondemolition 
(QND) technique can be used to avoid the detector backaction and to 
overcome the limitation on the signal-to-noise ratio. This work 
suggests such a QND technique and develops its quantitative 
description. 

QND measurement technique was proposed first for detection of 
weak forces acting on a harmonic oscillator in the context of the 
gravitational-wave antennas \cite{b9,b10}, and was discussed until 
now in application to measurements of various realizations of 
harmonic oscillators -- see, e.g., \cite{b11,b12}. Here the concept 
of a QND measurement is extended to the two-state system. In 
general, the QND measurement is realized when a quantum system is 
coupled to a measuring detector through an operator (the measured 
observable) that represents at least an approximate integral of 
motion. In this case, the backaction by the detector which increases 
fluctuations and uncertainty in the variables not commuting with the 
measured observable does not couple back into its evolution. Such a 
``decoupling'' of backaction makes it possible to measure the system 
continuously without significantly perturbing it. 

This discussion 
implies that specific scheme of the QND measurement of a two-state 
system should depend on the main part of the system Hamiltonian. In 
what follows, we consider the case of unbiased two-state system  
that is the most advantageous for the quantum coherent oscillations. 
In the basis of the two states of the oscillating variable $\hat{x} 
\propto \sigma_z$ (e.g., charge or flux states in the case of charge 
\cite{b3,b4} or flux \cite{b5,b6,b7} qubits, respectively) the 
Hamiltonian of the system is then  
\begin{equation} 
H=-\frac{1}{2} \Delta \sigma_x \, . 
\label{1} \end{equation} 
Here and below $\sigma$'s denote Pauli matrices, and $-\Delta/2$ 
is the amplitude of tunnel coupling between the basis states. 
The basic idea behind the QND measurement of such a system is 
illustrated in Fig. 1. In the spin-$1/2$ representation of the 
two-state system (\ref{1}), its dynamics can be seen as rotation 
with frequency $\Delta$ in the $y-z$ plane. To perform the QND 
measurement, the direction along which the spin is measured should 
follow as closely as possible the system rotation. This is 
achieved if the measurement direction rotates with frequency 
$\Omega \simeq \Delta$. (Since the phase of the oscillation of a 
spin $1/2$ can not maintain any semiclassical dynamics, there is 
no question of the phase relation between the two rotations.) 
Thus, one can suggest the following Hamiltonian for the QND qubit 
measurement:  
\begin{equation} 
H=-\frac{1}{2}\Delta \sigma_x -\frac{1}{2}(\cos \Omega t \sigma_z+ 
\sin \Omega t \sigma_y)f+ H_0\, . 
\label{2} \end{equation} 
Here $H_0$ is the Hamiltonian of the detector which is coupled to 
the qubit via the force $f$, and to simplify notations all coupling 
constants are included in the definition of $f$. 

\begin{figure}[htb]
\setlength{\unitlength}{1.0in}
\begin{picture}(3.,2.0) 
\put(.5,.0){\epsfxsize=1.9in\epsfbox{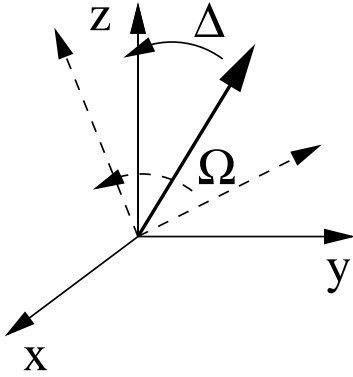}}
\end{picture}
\caption{Spin representation of the QND measurement of the quantum 
coherent oscillations of a qubit. The oscillations are represented 
as a spin rotation in the $z-y$ plane with frequency $\Delta$. 
QND measurement is realized if the measurement frame (dashed lines) 
rotates with frequency $\Omega \simeq \Delta$. }
\end{figure} 

The Hamiltonian (\ref{2}) of the QND qubit measurement is 
different from that of the non-QND measurement studied so far in 
theory \cite{b1,b2,b8,b13,b14,b15,b16} and used in experiments 
\cite{b3,b4,b5,b6} in the form of the qubit-detector coupling. 
The coupling term in (\ref{2}) should be contrasted with 
$-\sigma_z f/2$ in the non-QND case when the detector is coupled 
directly to the oscillating variable. Apart from this difference, 
the two situations should be similar, and the detector properties 
in (\ref{2}) can be taken to be the same as in the previously 
studied non-QND case: the detector is linear, with 
frequency-independent (in the frequency range given by the tunnel 
amplitude $\Delta$) response coefficient $\lambda$. These 
assumptions imply that the force $f$ can be viewed as random 
classical $\delta$-correlated variable with the correlation 
function  
\begin{equation} 
\langle f(t+\tau)f(t) \rangle_0 = 2\pi S_f \delta(\tau)\, ,   
\label{3} \end{equation}
where the average $\langle \ldots \rangle_0$ is taken over the 
detector density matrix, and $S_f$ is the constant low-frequency part 
of the spectral density of $f$, i.e., the detector backaction noise. 
For more detailed discussion of this detector model see \cite{b13}. 
Under the same assumptions, the correlation function of the 
detector response to the oscillations is: 
\begin{equation} 
K(\tau)= \frac{\lambda^2}{8} [\langle c(t)c(t+\tau)+c(t+\tau)c(t) 
\rangle ] \, ,  
\label{4} \end{equation}
where $c\equiv \cos \Omega t \sigma_z+ \sin \Omega t \sigma_y$ is 
the operator of qubit-detector coupling and the average is now 
taken both over the detector and qubit density matrices. The time 
dependence of $c$'s in eq.\ (\ref{4}) combines explicit time 
dependence in their definition and time evolution with the 
Hamiltonian (\ref{2}).  

To calculate the correlator (\ref{4}) we notice that the 
explicit time dependence of the coupling operator $c$ can be 
written as 
\begin{equation}
c=e^{i\Omega \sigma_x t/2} \sigma_z e^{-i\Omega \sigma_x t/2} \, . 
\label{5} \end{equation} 
This relation follows directly from properties of the Pauli matrices 
and expresses quantitatively the notion of ``rotation of the 
measurement direction'' in Fig. 1. Using this relation one can check 
that the time evolution operator $S$ associated with the Hamiltonian 
(\ref{2}) has a simple form in the rotating measurement frame: 
\[ S(t_1,t_2)=T\exp \left\{-i\int^{t_1}_{t_2}d t'H(t') \right\} \]

\vspace*{-2ex} 

\begin{equation} 
= e^{i\Omega \sigma_x (t_1-t_2)/2} e^{-iH'(t_1-t_2)} \, ,
\label{6} \end{equation}
where $H'$ is an effective Hamiltonian of the system in the rotating 
frame: 
\begin{equation} 
H'=-\frac{1}{2} (\Delta-\Omega) \sigma_x -\frac{1}{2}\sigma_z f+ 
H_0\, . 
\label{7} \end{equation} 

Using the fact (demonstrated more explicitly below) that the
correlator (\ref{4}) should be independent of the initial density 
matrix $\rho$ of the qubit, we can take $\rho$ in the simplest form 
$\rho=1/2$. Equations (\ref{5}) and (\ref{6}) allow us then to 
reduce the correlator (\ref{4}) to the following form: 
\begin{equation} 
K(\tau)=\frac{\lambda^2}{4} \mbox{Re} \langle \sigma_z 
\sigma_z(\tau) \rangle \, , \;\;\; 
\sigma_z(\tau) = e^{iH'\tau}\sigma_z e^{-iH'\tau} \, .
\label{8} \end{equation}
To find the average of the operator $\sigma (\tau)$ in (\ref{8}) over 
the detector backaction noise $f$, it is convenient to start with 
the Heisenberg equations for $\sigma_z(\tau)$ with the Hamiltonian 
$H'$. Averaging the resulting equations for the matrix elements 
$\sigma_{ij}$ of $\sigma_z (\tau)$ with the help of the correlator 
(\ref{3}) we get: 
\begin{equation}
\dot{\sigma}_{11}=  i\delta (\sigma_{12}-\sigma_{21})/2\, , \;\;\;\; 
\dot{\sigma}_{12}= i\delta (\sigma_{11}- \sigma_{22})/2 - 
\Gamma \sigma_{12}  \, ,  
\label{9} \end{equation}
and $\dot{\sigma}_{22}=-\dot{\sigma}_{11}$. Here $\delta \equiv 
\Delta-\Omega$, and $\Gamma=\pi S_f$ is the rate of backaction 
dephasing of the oscillations by the detector. 

As the next step, we need to take into account environment-induced 
energy relaxation/dephasing that affects the qubit in addition 
to the detector backaction. Assuming that interaction with 
the environment is weak, so that the characteristic relaxation rate 
is much smaller than the tunnel amplitude $\Delta$, we can simply 
add the corresponding terms in the equation (\ref{9}) for 
$\sigma_z (\tau)$. It is more convenient to do this not directly 
in the basis of eigenstates of the oscillating coordinate used in 
(\ref{9}), but in the basis of energy eigenstates of the qubit. 
Transforming eq.\ (\ref{9}) into the energy basis (in the spin 
notations, the transformation is the $\pi/2$ rotation around the 
$y$ axis: $\sigma_z \rightarrow  \sigma_x \, ,\sigma_x 
\rightarrow -\sigma_z$) and adding the term responsible for the 
environmental relaxation, we obtain the following equation for 
the off-diagonal matrix element of the operator $s$ of the 
oscillating variable (given by $\sigma_z(\tau)$ in the original 
coordinate basis):  
\begin{equation}
\dot{s}_{12}=i\delta s_{12} - \Gamma (s_{12}-s_{21})/2 - 
\Gamma_e s_{12} \, . 
\label{10} \end{equation}

To find the correlator (\ref{8}) we need to solve the evolution 
equations for $s$ with the initial conditions $s_{12}=1$ and the 
diagonal elements of $s$ equal to zero. Equations for the diagonal 
elements show then that they remain zero at all $\tau$, and 
$s_{12}(\tau)$ (\ref{10}) determines the correlator completely. 
This means that the real part of the product $ss(\tau)$ is 
proportional to the unity matrix, and the correlator (\ref{8}) 
is independent of initial qubit density matrix. We also note that 
the different form of the terms describing detector- and 
environment-induced 
dephasing in eq.\ (\ref{10}) is due to the QND nature of the 
detector-qubit coupling, and that while the environmental 
relaxation rate $\Gamma_e$ was assumed to be  much smaller than 
$\Delta$, both $\Gamma_e$ and $\Gamma$ can be larger than 
the detuning $\delta$. 

Solving eq.\ (\ref{10}) with the initial condition $s_{12}=1$ and 
substituting the solution in (\ref{8}) we obtain the correlation 
function of the detector response to the qubit: 
\begin{equation}
K(\tau)=\frac{\lambda^2}{4} e^{-(\Gamma/2 +\Gamma_e)\tau} 
\left[ \mbox{cosh} Dt +\frac{\Gamma}{2D}\mbox{sinh}   Dt \right] 
\, ,  
\label{11} \end{equation}
where $D\equiv \left( \Gamma^2/4-\delta^2 \right)^{1/2}$. 
The correlation function (\ref{11}) determines the spectral 
density $(1/\pi) \int d\tau K(\tau) \cos \omega \tau $ of the 
detector output. Including the constant output noise $S_q$ of the 
detector, the spectral density is: 
\begin{equation}
S(\omega ) = S_q +\frac{\lambda^2}{4\pi} 
\frac{(\Gamma +\Gamma_e)(\Gamma^2 +\Gamma \Gamma_e +\delta^2)  
+\Gamma_e \omega^2}{(\omega^2-\delta^2+\Gamma^2 +\Gamma 
\Gamma_e)^2+\omega^2(\Gamma+2\Gamma_e)^2} \, . 
\label{12} \end{equation}

If the detuning $\delta$ is much larger than the dephasing rates, 
the spectral density reduces to the two Lorentzian peaks at 
$\omega =\pm \delta$. In the vicinity of the positive-frequency 
peak, the spectral density can be written as   
\begin{equation}
S(\omega ) = S_q +\frac{\lambda^2}{8\pi} \frac{\Gamma_e +\Gamma/2}{ 
(\omega-\delta)^2 + (\Gamma_e +\Gamma/2)^2 } \, . 
\label{13} \end{equation}
This expression shows that for large detuning, the spectral density 
is close to the one obtained in the usual, non-QND measurement 
\cite{b8,b13}. The only difference is the frequency shift of the 
spectral peak from intrinsic oscillation frequency $\Delta$ by 
the frequency $\Omega$ of rotation of the measurement frame. The 
main goal of the QND technique, avoiding the detector backaction, is 
not reached in this regime. The backaction dephasing broadens the 
oscillation spectral line, and limits the height of the oscillation 
peak relative to the background set by the detector output noise. 
Indeed, eq.\ (\ref{13}) shows that even for $\Gamma_e=0$, the 
maximum height $S_{max}$ of the oscillation peak is 
$S_{max}=\lambda^2/4\pi \Gamma$, and the limitation on the response 
coefficient of the detector from the linear-response theory, 
$\lambda \leq 4\pi S_f S_q$ \cite{b13}, shows that the peak is 
limited as in the non-QND measurement: $S_{max}/S_q\leq 4$. 

The situation, however, changes, if the frequency of rotation 
of the measurement frame matches the oscillation frequency more 
accurately, so that $\delta \ll \Gamma$. The oscillation line is 
shifted then to zero frequency, and the lineshape is: 
\begin{equation}
S(\omega ) = S_q +\frac{\lambda^2}{2\pi} \frac{\Gamma_e +\gamma}{ 
\omega^2 + (\Gamma_e +\gamma)^2 } \, ,  \;\;\;\; 
\gamma \equiv \delta^2/\Gamma. 
\label{14} \end{equation}
Qualitatively, $\gamma$ in this equation is the rate of rare jumps 
of spin in the rotating measurement frame between the positive and 
negative measurement direction, and we see that the spectral line 
is now broadened not directly by the backaction dephasing but by 
these rare jumps. The most important feature of these jumps is that 
the rate $\gamma$ vanishes together with the detuning $\delta$, and 
the spectrum (\ref{14}) of the detector output 
reproduces then intrinsic linewidth of the oscillation unaffected by 
the backaction dephasing. Therefore, ``rotating'' measurement (\ref{2}) 
with frequency $\Omega$ equal to the tunnel frequency $\Delta$ avoids 
the detector backaction and realizes the QND measurement of the quantum 
coherent oscillations in a two-state system. Other methods of 
avoiding the backaction dephasing are based on control of the 
oscillations via the feedback \cite{b17}. These methods, however, do 
not represent real ``measurement'' but rather creation of the 
oscillations.  

The QND technique is required to measure the intrinsic spectral line 
of the oscillations in a qubit. Such an intrinsic oscillation line 
is typically found in calculations that implicitly assume that the 
spectral density can be measured without disturbance from the 
detector even in quantum mechanics -- see, e.g., \cite{b18}. 
As we saw above, in the case of mesoscopic qubits this assumption 
is not obvious, and non-trivial measurement schemes are required to 
observe intrinsic oscillation spectrum. Without these schemes, the 
detector backaction affects the spectrum at least within the linear 
detection approach appropriate for typical mesoscopic detectors.  

In mesoscopic qubits, for which the question of measurement of an 
individual qubit is particularly relevant, the necessity to measure
two non-commuting qubit operators to implement the QND measurement 
presents a non-trivial requirement since only one (basis-forming) 
observable typically has clear physical realization. A system 
where this requirement can be satisfied is the 
Josephson-junction qubit that combines charge and flux dynamics. 
The simplest version of such a qubit is the ``charge-controlled 
SQUID'': two junctions included in a superconducting loop with 
inductance $L$, with the island between the junctions that has 
small electric capacitance $C$ (Fig.\ 2) \cite{b19}. External 
magnetic flux $\Phi_e$ is applied to the loop, and a gate 
electrode induces charge $q$ on the middle island. 

\begin{figure}[htb]
\setlength{\unitlength}{1.0in}
\begin{picture}(3.5,1.25) 
\put(.0,.1){\epsfxsize=3.2in \epsfbox{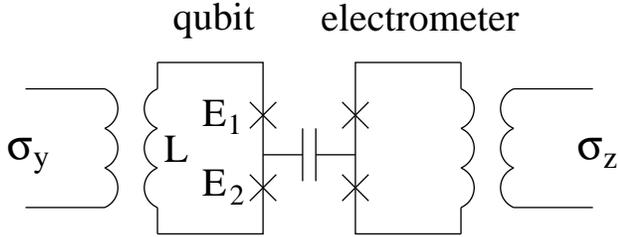}}
\end{picture}
\caption{Schematic of the Josephson-junction qubit structure 
that enables measurements of the two non-commuting observables 
of the qubit, $\sigma_z$ and $\sigma_y$, as required in the 
QND Hamiltonian (\ref{2}). For discussion see text. }
\end{figure} 
Depending on the parameter values, this system can act as a charge 
or flux qubit. To be specific, we consider here the situation when 
the junction coupling energies $E_{1,2}$ are small in comparison to 
the island charging energy $(2e)^2/2C$, and the inductance $L$ is 
also small. Then, dynamics of charge $2en$ on the middle island 
dominates the system, and for $q\simeq e$ is reduced to the 
two-state dynamics with the basis states $n=0$ and $n=1$. If the 
frequency of the ``plasma'' oscillations of the flux $\Phi$ 
in the loop $L$ around the external flux $\Phi_e$ is much 
larger than the energies of the charge qubit, the oscillations 
effectively decouple from the charge dynamics, and the charge part 
of the system Hamiltonian reduces then to: 
\begin{equation}
H=\frac{e(e-q)}{C} \sigma_z - E_+ \cos \varphi_e \sigma_x + 
E_- \sin \varphi_e \sigma_y\, .  
\label{16} \end{equation}
Here $E_{\pm}\equiv (E_1\pm E_2)/2$, Pauli matrices act in the 
basis of the two charge states $n=0$ and $n=1$, and $\varphi_e 
\equiv e\Phi_e/\hbar$. 

The flux part of the system still plays important role, since 
the tunneling of Cooper pairs in the charge qubit (\ref{16}) 
produces current in the loop: 
\begin{equation}
I= \frac{1}{2} (I_+ \sin \varphi_e \sigma_x + I_-  \cos \varphi_e 
\sigma_y ) \, , 
\label{17} \end{equation}
where $I_j=2e E_j /\hbar$ are the junction critical currents, and 
$I_{\pm}\equiv (I_1\pm I_2)/2$. (The last equation assumes, for 
simplicity, that the junction capacitances are equal.) For finite 
loop inductance $L$, the 
current (\ref{17}) creates small variations of the flux $\Phi$ 
through the loop that can be detected by external system, and by 
monitoring these variations one can measure either $\sigma_x$ or 
$\sigma_y$ component of the charge qubit (\ref{16}). The 
possibility to measure all observables of an individual mesoscopic 
qubit provided by the qubit (\ref{16}) can be important for many 
different purposes.

When $\Phi_e=0$ and $q=e$, the qubit (\ref{16}) provides the 
necessary elements for the QND measurement discussed above. In 
this case, eq.\ (\ref{16}) reduces to the unbiased Hamiltonian 
(\ref{1}) with 
the tunnel amplitude $E_+$, and the current $I=(I_-/2) \sigma_y$ 
in the inductance $L$ represents the $\sigma_y$ component of 
the qubit dynamics. The $\sigma_z$ component can obviously be 
measured through the charge on the middle electrode of the qubit. 
The final step in realization of the QND Hamiltonian (\ref{2}) 
is to convert both measurement components into one physical form 
(e.g., charge or flux) and apply (with sine- and cosine-modulated 
coupling strength) to one detector. One way of achieving this 
is to convert the charge signal ($\sigma_z$) into the flux form
by another charge-controlled SQUID operated in the 
regime of Bloch electrometer \cite{b20} (Fig.\ 2), when all 
internal frequencies of the electrometer are much larger than 
the frequency $E_+$ of the charge signal. The signal modulates 
then the quasistationary critical current of the double junction 
system and, as a result, changes the current and the flux in 
the electrometer loop. Once the two signals are in the flux form, 
the subsequent steps can be realized using known 
Josephson-junction circuits: modulated flux transformers and 
magnetometer. 

To summarize, it is possible to design the QND technique for 
measurement of the quantum coherent oscillations in an 
individual two-state system. The technique avoids the detector 
backaction, and overcomes the limitation on the signal-to-noise 
ratio of the measurement of the spectral density of the 
oscillations imposed by the backaction dephasing.

\vspace*{.4ex} 

This work was supported in part by the AFOSR, and by the NSA 
and ARDA under the ARO contract. The author would like to thank 
K.K. Likharev and V.K. Semenov for useful discussions.


\vspace*{3ex}

\begin{references} 

\bibitem{b1} D.V. Averin, Fortschrit. der Physik {\bf 48}, 1055 (2000). 

\bibitem{b2} Yu. Makhlin, G. Sch\"{o}n, and A. Shnirman, Rev.\ Mod.\ Phys. 
{\bf 73}, 357 (2001).

\bibitem{b3} Y. Nakamura, Yu.A. Pashkin, and J.S. Tsai, Nature {\bf 398}, 
786 (1999). 

\bibitem{b4} V. Bouchiat, D. Vion, P. Joyez, D. Esteve, and M.H. Devoret, 
J. Supercond. {\bf 12}, 789 (1999).  

\bibitem{b5} J.R. Friedman, V. Patel, W. Chen, S.K. Tolpygo, and J.E.
Lukens, Nature {\bf 406}, 43 (2000).

\bibitem{b6} C.H. van der Wal, A.C.J. ter Haar, F.K. Wilhelm, R.N.
Schouten, C. Harmans, T.P. Orlando, S. Lloyd, and J.E. Mooij, 
Science {\bf 290}, 773 (2000).

\bibitem{b7} S.Y. Han, Y. Yu, X. Chu, S.I. Chu, and Z. Wang, 
Science {\bf 293}, 1457 (2001). 

\bibitem{b8} A.N. Korotkov and D.V. Averin, Phys.\ Rev. B {\bf 64}, 
165310 (2001).

\bibitem{b9} C.M. Caves, K.S. Thorne, R.W.P. Drever, V.D. Sandberg, 
and M. Zimmermann, Rev.\ Mod.\ Phys. {\bf 52}, 341 (1980).

\bibitem{b10} V.B. Braginsky and F.Ya. Khalili, Rev.\ Mod.\ Phys. 
{\bf 68}, 1 (1996).

\bibitem{b11}  P. Grangier, J. A. Levenson, and J.-P. Poizat, Nature 
{\bf 396}, 537 (1998).
 
\bibitem{b12} S. Peil and G. Gabrielse, Phys.\ Rev.\ Lett. {\bf 83}, 
1287 (1999).

\bibitem{b13} D.V. Averin, in: {\em ``Exploring the Quantum-Classical 
Frontier: Recent Advances in Macroscopic and Mesoscopic Quantum 
Phenomena''}, Eds. J.R. Friedman and S. Han, to be published; 
cond-mat/0004364. 

\bibitem{b14} Yu. Makhlin, G. Sch\"{o}n, and A. Shnirman, 
Phys.\ Rev.\ Lett. {\bf 85}, 4578 (2000).

\bibitem{b15} M.H. Devoret and R.J. Schoelkopf, Nature 
{\bf 406}, 1039 (2000).

\bibitem{b16} A. Aassime, G. Johansson, G. Wendin, R.J. Schoelkopf, 
and P. Delsing, Phys.\ Rev.\ Lett. {\bf 86}, 3376 (2001).   

\bibitem{b17} R. Ruskov and A.N. Korotkov, cond-mat/0107280. 

\bibitem{b18} M.-S. Choi, F. Plastina, and R. Fazio, Phys.\ Rev.\ 
Lett. {\bf 87}, 116601 (2001).  

\bibitem{b19} J.R. Friedman and D.V. Averin, Phys.\ Rev.\ Lett. 
{\bf 88}, 050403 (2002).   

\bibitem{b20} A.B. Zorin, Phys.\ Rev.\ Lett. {\bf 86}, 3388 (2001). 

\end{references}
\end{document}